**Fantastic Behavior of High-$T_C$ Superconductor Junctions: Tunable Superconductivity**

CHEN Yuan-Sha (陈沅沙), Zhang Yan (张炎),

LIAN Gui-Jun (连贵君), and XIONG Guang-Cheng *(熊光成)

Department of Physics

Peking University, 100871 Beijing, P. R. China

Carrier injection performed in oxygen-deficient $YBa_2Cu_3O_{7-\delta}$ (YBCO) hetero-structure junctions exhibited tunable resistance that was entirely different with behaviors of semiconductor devices. Tunable superconductivity in YBCO junctions, increasing over 20 K in transition temperature, has achieved by using electric processes. To our knowledge, this is the first observation that intrinsic property of high $T_C$ superconductors' superconductivity can be adjusted as tunable functional parameters of devices. The fantastic phenomenon caused by carrier injection was discussed based on a proposed charge carrier self-trapping model and BCS theory.

Development of materials has resulted in remarkable changes of our daily life. Transition metal oxides demonstrate several unexpected phenomena such as the highlight of the Nobel Prize-winning discovery of high-temperature superconductors. Recently, several surprise results were reported that resistance of single crystal $Pr_{0.7}Ca_{0.3}MnO_3$ (PCMO) dropped off over three orders of magnitude at 20 K in high electric voltages.[1] It was also reported that resistance of PCMO epitaxial films[2] and other poor conductive oxides[3-5] switched between high resistance states (HRS) and low resistance states (LRS) in electric fields as tunable resistance behavior, which may have potential application future in the next generation of nonvolatile memory. Although, the mechanism of tunable resistance for oxides is not clear at this moment, a following question should emerge as whether intrinsic properties of materials could be adjusted by electric process, which would be very interesting in fundamental researches and applications.

Carrier injection performed in $Pr_{0.7}Ca_{0.3}MnO_3$ (PCMO) hetero-structure junctions exhibited stable resistance switching between high resistance states (HRS) and low resistance states (LRS)[6]. Figure 1 shows room-temperature *I-V* data of a PCMO junction consisted of *n*-type oxide semiconductor of Nb(0.6%)-doped $SrTiO_3$ (Nb-STO)[7,8], where the doping concentration of Nb was 0.6 wt.%. In low current-voltage region, effective resistance of the PCMO device was higher than 1 MΩ with rectifying behavior. When increasing current, resistance of the device abruptly jumped to LRS of about 100 Ω. The LRS is stable without electric fields or even with low negative-voltage. The LRS of device can switch back to HRS in negative-current scan, which demonstrate a setting and resetting loop by electric processes. The nearly linear *I-V* character of LRS exhibits behavior of typical pure resistance, which indicates disappearance of interface barriers. Because the stable LRS was induced by charge carrier injection, the observation exhibited that the injected carriers should still stayed in the devices [6] and resulted in dramatic change in resistance and interface bias-barriers.

High-$T_C$ oxide superconductors, such as $YBa_2Cu_3O_7$ (YBCO), show superconducting transition temperatures of $T_C$ with two fundamental characteristics of superconductivity, the disappearance of resistance and the expulsion of magnetic flux that is so called Meissner effect [9]. Superconducting transition temperatures of samples can be measured by resistance and magnetic measurements receiving $T_{CR}$ and $T_{CM}$. It is believed that $T_C$ demonstrates intrinsic property of the carriers in high-$T_C$ materials, for example, the electron band structure and carrier density. It is known that for oxygen-deficient YBCO, $T_C$ depends on oxygen compositions of 7-δ in samples [10]. Figure 2 shows data of field-cooling (FC) and zero-field-cooling (ZFC) moments measured by a SQUID system for an oxygen-deficient YBCO films that was deposited by pulsed laser deposition technique at 780℃ in 2 Pa of oxygen and cooled to room temperature in same environment. In measurements, ZFC signal exhibit screening effect of surface and grain boundary. Not distinct signal of diamagnetic moment was observed down to 8 K, which indicated no Meissner effect until 8K and $T_C$ < 8 K. According to experimental data reported by Jorgensen *et al.* [10], δ for our oxygen-deficient YBCO film should be greater than 0.65. Otherwise some oxygen atoms may be attracted on surface of the oxygen-deficient YBCO film in atmosphere condition at room temperature.

By using pulsed laser deposition technique, insulating STO buffer layers(about 200nm thick) were deposited on Nb-STO substrates at 800℃ in 2 Pa of oxygen and cooled to room temperature in 0.5 atm of oxygen. Then oxygen deficient YBCO films were deposited on substrates, which were partially on insulating STO buffer layers and partially on Nb-STO substrate (with geometry that indicated in the inset of Fig. 3). Figure 3 shows room-temperature *I-V* data of a YBCO/Nb-STO junction, of which the effective resistance changes between HRS and LRS as that observed in PCMO/Nb-STO junctions.

Figure 4(a) shows a series of FC data of one oxygen-deficient YBCO/Nb-STO junction with geometry indicated in the inset of Fig. 3. When the oxygen deficient YBCO/Nb-STO junction was in HRS state (about 20-30 kΩ), no distinct signal of Meissner effect was observed in data A that indicated $T_C$ < 8 K for this oxygen deficient YBCO film. Figure 4(b) shows *I-V* data of the oxygen-deficient YBCO/Nb-STO junction. With increased current, the junction's HRS (about 200 kΩ) was adjusted to LRS. Inset of Fig. 4(b) is *R-V* data of the last current scan with LRS of about 10 (50-5x10$^{-4}$ mA) to 250 Ω (2x10$^{-5}$ mA ). It is very interesting that when keeping with such LRS state for a magnetic measurement, FC data B of the YBCO/Nb-STO junction demonstrated a dramatic drop with diamagnetic moment (2.2x10$^{-6}$ emu at 31.87 K) as Meissner effect at $T_C$ of 34.81 K. This is an important observation because change of the superconducting transition temperature $T_C$ from less than 8 K to 34.81 K (at least more than 20 K) is obtained in one and the same sample, which is only operated by electric process. We further adjusted the junction by negative-current scan but cannot recover rectifying behavior for the junction. However, FC data C indicated that $T_C$ of the oxygen deficient YBCO film dropped to 23.59 K (2.0x10$^{-6}$ emu at 20.84 K) after this electric processes. In these processes, Figure 4 clearly demonstrates that superconductivity of one oxygen-deficient YBCO film was adjusted by electric processes. Similar results were obtained in other YBCO/Nb-STO junctions that electric processes can adjust superconductivity of samples.

After the discovery of high temperature superconductor, efforts to increase $T_C$ rely on chemical processes, for example, from $La_{2-x}Sr_xCuO_4$ with $T_C$ about 40 K to $YBa_2Cu_3O_7$ with $T_C$ about 90 K and then to Bi, Tl and Hg compounds with $T_C$ over 100 K [9]. Our obtained phenomenon should be the first observation, to our knowledge, that intrinsic property of high $T_C$ superconductors, superconductivity can be modified by using electric processes as tunable functional parameters of devices. The tunable superconductivity obtained from carrier injection could be understood by considering the charge carrier self-trapping model [6] with the BCS theory.

From carrier injection in PCMO/Nb-STO junctions, tunable resistance and disappearance of interface bias-barriers indicate the influence of the injected carriers, which should be related to effects of self-trapping carriers [6]. Although PCMO has strongly correlated electrons, the tunable resistance behavior exhibits importance of the self-trapping carriers in conduction. As described in BCS theory, weak electron-phonon coupling contributes virtual phonon for electron-electron interaction that results in Cooper pairs and superconductivity. Since carrier injection in oxygen-deficient YBCO junctions would introduce self-trapping carriers as minority carriers. With suitable electron-phonon coupling under actual periodic lattices of YBCO, it is reasonable that superconducting transition of oxygen-deficient YBCO was achieved and modified by changes of the self-trapping carriers. Carrier injection in oxide electronics devices has introduced tunable superconductivity and resistance by the self-trapping carriers that open a novel area for searching new oxide electronics devices with other fantastic functions and novel applications.

We acknowledge supporting from National Natural Science Foundation of China and helpful discussions with Z. Z. Gan, Z. X. Zhao and Y. H. Zhang.

* To whom correspondence should be addressed. Email: ssxgc@pku.edu.cn

**Figure Captions**

Figure 1 (color). *I-V* data of a PCMO/Nb-STO junction.

Figure 2 (color). Data of field-cooked (FC) and zero-field-cooled (ZFC) moments for an oxygen-deficient $YBa_2Cu_3O_{7-\delta}$ film measured by a SQUID system.

Figure 3 (color). Room temperature *I-V* data of an oxygen-deficient YBCO/Nb(0.6%)-STO junction. Inset is schematic drawing of the junction.

Figure 4 (color). (a) FC data of one oxygen-deficient YBCO/Nb(0.6%)-STO junction before and after electric processes. FC data A (black ▼) without Meissner effect were that the junction was in HRS before electric process. FC data B (red ●) were that the junction was in LRS after electric process, in which a dramatic drop with diamagnetic moment ($2.2 \times 10^{-6}$ emu at 31.87 K) was due to Meissner effect at $T_C$ of 34.81 K. After further adjusted the junction by electric process, FC data C (green ▲) exhibited $T_C$ of 23.59 K. (b) *I-V* data of the YBCO/Nb-STO junction in HRS and LRS. Inset of (b) is *R-V* data of the YBCO/Nb-STO junction that demonstrated the change from HRS (about 200 kΩ) to LRS of about 10 ($50-5 \times 10^{-4}$ mA) to 250 Ω ($2 \times 10^{-5}$ mA). The LRS in Fig. 4(b) was the situation for FC measurement of data B in Fig. 4(a).

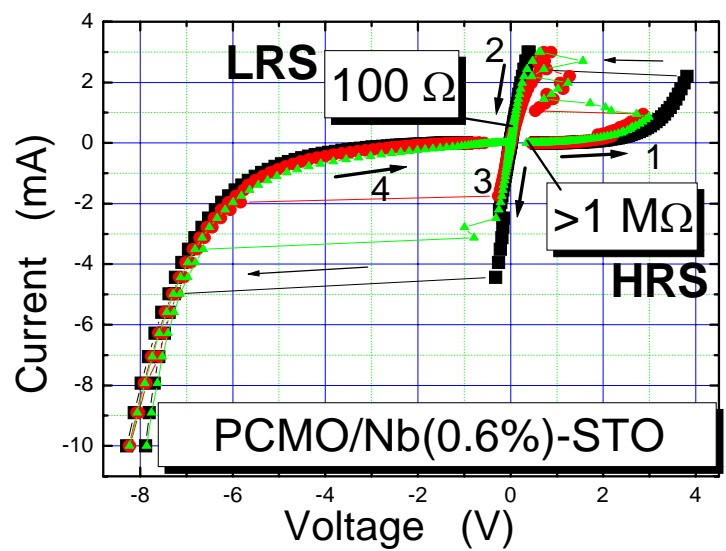

Figure 1

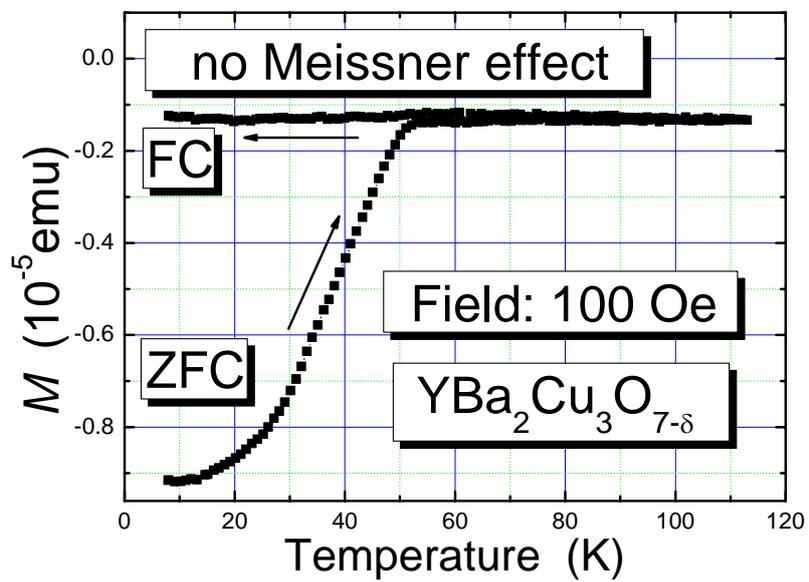

Figure 2

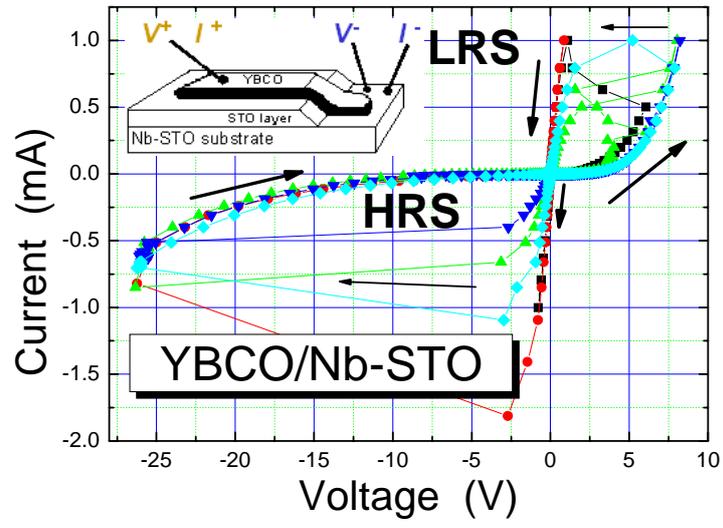

Figure 3

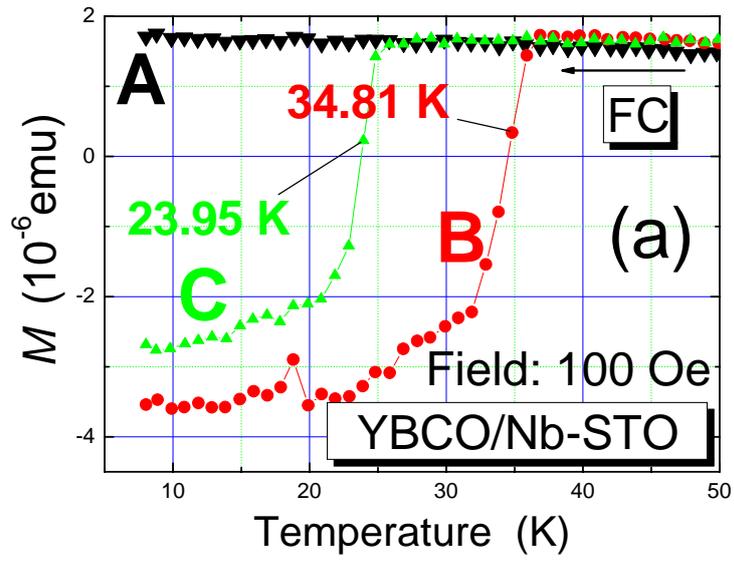

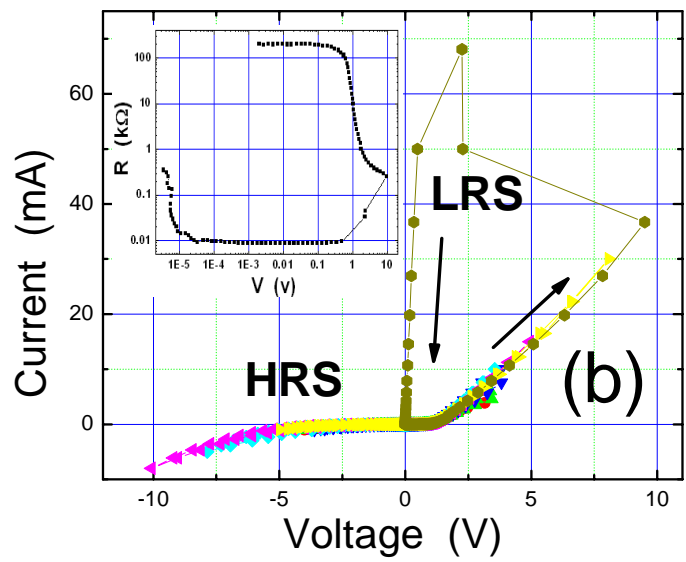

Figure 4